\documentclass[pdflatex,iicol,sn-basic]{sn-jnl}

\usepackage[T1]{fontenc}
\usepackage[utf8]{inputenc}
\usepackage{microtype}
\usepackage{graphicx}
\usepackage{amsmath,amssymb,amsfonts,mathtools,bm}
\usepackage{multirow}
\usepackage{booktabs}
\usepackage{flushend}
\usepackage{rotating}
\usepackage{float}
\usepackage{algorithm}
\usepackage{algorithmic}
\usepackage{enumitem}
\hypersetup{
  hypertexnames=false,
  pdftitle={Spatial-sign-based multilinear principal component analysis for tensor data},
  pdfauthor={Dongxu Yang, Wanfeng Liang, Le Zhou and Long Feng},
  pdfkeywords={elliptical distribution, multilinear principal component analysis, robust dimension reduction, spatial median, spatial sign, tensor data}
}

\theoremstyle{thmstyleone}
\newtheorem{theorem}{Theorem}[section]
\newtheorem{proposition}[theorem]{Proposition}
\newtheorem{corollary}[theorem]{Corollary}

\newtheorem{assumption}[theorem]{Assumption}
\theoremstyle{thmstyletwo}

\newcommand{\E}{\mathbb{E}}
\newcommand{\Prob}{\mathbb{P}}
\newcommand{\tr}{\operatorname{tr}}
\newcommand{\diag}{\operatorname{diag}}

\newcommand{\op}{\mathrm{op}}
\newcommand{\vecc}{\operatorname{vec}}
\newcommand{\unvec}{\operatorname{unvec}}

\begin{document}

\makeatletter
\newcommand{\smpcapreservedlabel}[2]{%
  \expandafter\gdef\csname r@#1\endcsname{{#2}{}{}{}{}}}
\smpcapreservedlabel{sec:auxiliary}{A}
\smpcapreservedlabel{sec:covered-distributions}{A.3}
\smpcapreservedlabel{thm:population-structure}{A.1}
\smpcapreservedlabel{prop:projector-geometry}{A.2}
\smpcapreservedlabel{prop:empirical-gaps}{A.3}
\smpcapreservedlabel{prop:sample-solutions}{A.4}
\smpcapreservedlabel{prop:reconstruction-details}{A.5}
\makeatother

\title[Spatial-sign-based multilinear PCA]{Spatial-sign-based multilinear principal component analysis for tensor data}

\author[1]{\fnm{Dongxu} \sur{Yang}}
\author[2]{\fnm{Wanfeng} \sur{Liang}}
\author*[3]{\fnm{Le} \sur{Zhou}}\email{lezhou@hkbu.edu.hk}
\author*[1]{\fnm{Long} \sur{Feng}}\email{flnankai@nankai.edu.cn}

\affil*[1]{\orgdiv{School of Statistics and Data Science}, \orgname{Nankai University}, \orgaddress{\street{94 Weijin Road}, \city{Tianjin}, \postcode{300071}, \country{China}}}
\affil[2]{\orgdiv{School of Data Science and Artificial Intelligence}, \orgname{Dongbei University of Finance and Economics}, \orgaddress{\street{217 Jianshan Street}, \city{Dalian}, \postcode{116025}, \country{China}}}
\affil[3]{\orgdiv{Department of Mathematics}, \orgname{Hong Kong Baptist University}, \orgaddress{\street{224 Waterloo Road}, \city{Hong Kong}, \country{China}}}

\abstract{Multilinear principal component analysis (MPCA) reduces the dimension of tensor-valued data while preserving their mode-specific structure, but its quadratic scatter criterion can be unstable under heavy-tailed distributions and contamination. We propose spatial-sign-based multilinear principal component analysis (SMPCA), a robust dimension-reduction method that centers the observations by their spatial median, removes radial magnitude through spatial-sign normalization, and estimates the mode-wise loading spaces by alternating eigendecompositions. Under a separable tensor elliptical model, we show that the target mode-wise loading spaces uniquely maximize the population criterion and that one complete sweep of exact population block updates recovers them from any initialization. We also characterize exactly when their tensor-product subspace coincides with a leading unrestricted subspace of vectorized spatial-sign PCA and, when finite second moments exist, ordinary vectorized PCA. At the sample level, we derive explicit statistical rates for the mode-wise subspaces and the joint multilinear projector, obtain corresponding reconstruction guarantees, establish consistency of the cumulative-contribution dimension selector, and prove that the objective values generated by exact cyclic updates are nondecreasing and convergent. Simulations and an empirical application show that SMPCA is more accurate and stable than competitors under heavy-tailed distributions and outlier contamination, while retaining competitive performance under light-tailed settings.}


\maketitle

\section{Introduction}
Principal component analysis (PCA) is a fundamental tool for dimension reduction, feature extraction and low-rank representation \citep{jolliffe2002principal}. In many modern applications, however, one observation is naturally a matrix or a higher-order tensor rather than a vector. Examples include images, videos, neuroimaging arrays, longitudinal panels and multivariate time series. Such data contain distinct structural modes, and vectorization can obscure their mode-specific interpretation and dependence structure while collapsing several moderate mode dimensions into a single, potentially very large product dimension \citep{kolda2009tensor,bi2021tensors,Chen2019FactorMF,Han2022}.

Classical multiway decompositions, including Tucker and PARAFAC decompositions, represent tensor data through low-dimensional mode-specific factors \citep{tucker1966some,Harshman1970FoundationsOT,Carroll_Chang_1970}. For matrix-valued observations, two-dimensional PCA estimates a loading space along one matrix mode, whereas two-directional two-dimensional PCA estimates loading spaces along both the row and column modes without vectorizing the data \citep{1261097,ZHANG2005224}. \citet{lu2008mpca} generalized this mode-wise approach to higher-order tensors through multilinear principal component analysis (MPCA), which seeks mode-wise projections that preserve as much total tensor scatter as possible. The resulting low-dimensional tensor representation preserves the multiway structure and mode-specific interpretation of the data.

A substantial literature has extended MPCA in different directions. Uncorrelated and non-negative variants impose additional structure on the extracted components \citep{5272374,5337979}; concurrent-subspace and Bayesian formulations adapt the multilinear representation to recognition and latent modeling tasks \citep{4358655,4633660}; and online algorithms update the loading spaces as new tensor observations arrive \citep{HAN2018888}. Multilinear projections have also been incorporated into neural architectures for tensor classification \citep{7867073}. In parallel, statistical work on tensor factor models, generalized low-rank tensor estimation and multiway principal components has developed structure-aware low-dimensional models and estimators with statistical or computational guarantees \citep{Chen2019FactorMF,Han2022,ouyang2025multiway}.
 
While existing literature has demonstrated the value of preserving tensor structure, most MPCA methods and their variants were developed based on empirical second moments, which can be unstable when the distribution is heavy-tailed or the sample contains outlying observations. This issue is especially relevant for tensor data: a single anomalous tensor can dominate a quadratic scatter criterion through its squared Frobenius norm, while contamination in only a few tensor entries can affect the scatter estimates for multiple modes. Existing robust extensions address such contamination in different ways. \citet{inoue2009robust} proposed separate iterative procedures for casewise and cellwise outliers. Other approaches replace the quadratic criterion with an $L_p$-type objective or develop block tensor PCA using projection criteria based on the Frobenius norm rather than its square \citep{7924765,ZHANG2024112712}. More recently, \citet{hirari2026robust} introduced a unified method that simultaneously handles casewise and cellwise outliers and accommodates missing entries; its objective is optimized by iteratively reweighted least squares. These methods broaden the applicability of MPCA, although their loading-space estimation procedures generally require repeated reweighting or nonquadratic optimization.

Spatial signs provide a different route to robust principal-subspace estimation. After robust centering, each observation is divided by its norm, so its contribution to the resulting spatial-sign covariance matrix depends on its direction but not on its distance from the center \citep{oja2010multivariate,taskinen2012robustifying}. For elliptically distributed vectors, the population spatial-sign covariance matrix has the same eigenvectors and eigenvalue ordering as the shape matrix, although its eigenvalues are nonlinear transformations of the shape spectrum \citep{durre2016eigenvalues}. Thus, spatial-sign PCA can target the same principal subspaces without requiring finite second moments. The effects of estimating the center and the asymptotic behavior of the spatial median are well understood in fixed dimension \citep{durre2014unknownlocation,mottonen2010spatialmedian}, and recent work establishes non-asymptotic and high-dimensional guarantees for spatial-sign PCA \citep{zhao2024spatialsignbasedprincipal}. Existing spatial-sign PCA methods, however, treat each observation as a vector and therefore do not directly estimate mode-wise tensor loading spaces or exploit their multilinear parameterization.

Motivated by this gap, in this work, we develop spatial-sign-based multilinear principal component analysis (SMPCA). SMPCA combines spatial-median centering and spatial-sign normalization with separate mode-wise loading-space estimation. Spatial-sign normalization limits the influence of observations with unusually large Frobenius norms, while the multilinear formulation retains mode-specific structure.

The contributions of this paper are threefold.
\begin{enumerate}[label=(\roman*),leftmargin=2.2em,itemsep=2pt,topsep=3pt]
\item We develop the robust SMPCA by combining spatial-median centering and spatial-sign normalization with alternating mode-wise updates. Each update reduces to an eigendecomposition of a mode-specific scatter matrix. As a result, the proposed algorithm is computationally efficient. Our method preserves the multilinear structure of the data while limiting the influence of outlying observations, thereby yielding robust estimates of the loading spaces.
\item We establish population and finite-sample theory for SMPCA under a separable tensor elliptical model. At the population level, we prove that the target mode-wise loading spaces uniquely maximize the population criterion and that one complete sweep of exact population updates recovers them from any initialization. We also characterize exactly when their tensor-product subspace coincides with a leading unrestricted rank-$K$ target of vectorized spatial-sign PCA and, when finite second moments exist, ordinary vectorized PCA. At the sample level, we derive explicit statistical error rates for the mode-wise subspaces, the joint multilinear projector, and the reconstruction map. We further prove consistency of the cumulative-contribution rule for selecting the mode dimensions and establish that the objective-value sequence generated by exact cyclic updates is nondecreasing and convergent.
\item We evaluate SMPCA in simulations under Gaussian, heavy-tailed, and mixture-contaminated distributions and in a contaminated face-image reconstruction study. The simulations show subspace-estimation error comparable to that of MPCA under Gaussian sampling and generally lower errors under heavy-tailed and mixture-contaminated sampling. The face-image study shows improved reconstruction under contamination. In the reported implementations, SMPCA also runs faster than ROMPCA and TPCA-\(L_p\).
\end{enumerate}

The remainder of the paper is organized as follows. Section~2 introduces tensor operations, spatial signs and the population model. Section~3 presents SMPCA and its implementation. Section~4 develops the theoretical properties, including the comparison with unrestricted vectorized targets. Sections~5 and~6 report the simulation and face-image analyses, respectively. Section~7 concludes. Auxiliary results and proofs are collected in the appendices.

\section{Preliminaries}
Throughout the paper, scalars are denoted by italic letters, vectors by bold lower-case letters, matrices by bold upper-case letters and tensors by calligraphic letters. For an $N$th-order tensor $\mathcal X\in\mathbb R^{P_1\times\cdots\times P_N}$, $\mathbf X_{(n)}$ denotes its mode-$n$ unfolding, $\vecc(\mathcal X)$ its vectorization under the reverse lexicographic convention specified below, and $\unvec(\mathbf x)$ the inverse vectorization into the stated tensor dimensions. The Frobenius, spectral and nuclear norms are denoted by $\|\cdot\|_F$, $\|\cdot\|_{\op}$ and $\|\cdot\|_*$, respectively. For a symmetric matrix $\mathbf A$, $\lambda_j(\mathbf A)$ denotes its $j$th largest eigenvalue; for positive-definite $\mathbf A$, write $\kappa(\mathbf A)=\lambda_{\max}(\mathbf A)/\lambda_{\min}(\mathbf A)$. For positive deterministic sequences, $a_M\asymp b_M$ means that $a_M/b_M$ is bounded above and away from zero. For a non-negative random sequence $X_M$, the notation $X_M=O_P(a_M)$ means that, for every $\eta>0$, a constant $C_\eta<\infty$ exists such that $\Prob(X_M>C_\eta a_M)<\eta$ for all sufficiently large $M$. Matrix and vector stochastic orders are understood after applying the norm displayed in the corresponding statement. Unless fixed dimensions are stated explicitly, stochastic orders are taken as $M\to\infty$ along the dimension sequence specified in the corresponding result. The notation $\mathbf 1\{A\}$ denotes the indicator that the statement $A$ holds. If $\mathbf V$ has orthonormal columns, $\mathbf P_{\mathbf V}=\mathbf V\mathbf V^T$ is its orthogonal projector. For two loading matrices with $r$ orthonormal columns, define
\[
d(\widehat{\mathbf V},\mathbf V)
=2^{-1/2}\|\mathbf P_{\widehat{\mathbf V}}-\mathbf P_{\mathbf V}\|_F.
\]
This distance depends only on the corresponding subspaces and is therefore invariant to the choice of orthonormal bases.
\subsection{Tensor operations}
A mode-$n$ fiber of an $N$th-order tensor $\mathcal X\in\mathbb R^{P_1\times\cdots\times P_N}$ is obtained by fixing every index except the $n$th. The mode-$n$ unfolding $\mathbf X_{(n)}\in\mathbb R^{P_n\times\prod_{j\ne n}P_j}$ arranges these fibers as columns in reverse lexicographic order. For $\mathbf B\in\mathbb R^{J\times P_n}$, the mode-$n$ product $\mathcal Y=\mathcal X\times_n\mathbf B$ is defined by
\[
y_{i_1\cdots j\cdots i_N}
=\sum_{k=1}^{P_n}x_{i_1\cdots k\cdots i_N}b_{jk}.
\]
Repeated mode products are written as $\mathcal X\times_1\mathbf B_1\cdots\times_N\mathbf B_N$. The vectorization identity consistent with our unfolding convention is
\begin{equation}\label{eq:vectorization-identity}
\vecc(\mathcal X\times_1\mathbf B_1\cdots\times_N\mathbf B_N)
=(\mathbf B_N\otimes\cdots\otimes\mathbf B_1)\vecc(\mathcal X).
\end{equation}
The tensor inner product and Frobenius norm are
\[
\begin{aligned}
\langle\mathcal X,\mathcal Y\rangle
&=\sum_{i_1,\ldots,i_N}x_{i_1\cdots i_N}y_{i_1\cdots i_N},\\
\|\mathcal X\|_F
&=\langle\mathcal X,\mathcal X\rangle^{1/2}.
\end{aligned}
\]
A Tucker representation has the form $\mathcal X=\mathcal G\times_1\mathbf B_1\cdots\times_N\mathbf B_N$, where $\mathcal G$ is a core tensor \citep{tucker1966some,kolda2009tensor}.

\subsection{Multilinear principal component analysis}
For centered observations $\{\mathcal X_m\}_{m=1}^M$, MPCA estimates matrices $\mathbf V_n\in\mathbb R^{P_n\times K_n}$ satisfying $\mathbf V_n^T\mathbf V_n=\mathbf I_{K_n}$ by maximizing
\begin{equation}\label{eq:mpca-objective}
\sum_{m=1}^M
\|\mathcal X_m\times_1\mathbf V_1^T\cdots\times_N\mathbf V_N^T\|_F^2.
\end{equation}
For fixed loading matrices in all but one mode, the criterion is a trace maximization problem whose solution is the leading eigenspace of a mode-wise scatter matrix. Cyclically repeating these conditional maximizations gives the usual alternating least-squares algorithm \citep{gabriel1978least,lu2008mpca}. Because each summand in \eqref{eq:mpca-objective} scales quadratically under radial rescaling, an observation with a large Frobenius norm can dominate the criterion. This sensitivity motivates the spatial-sign modification below.

\subsection{Spatial signs and the spatial median}
For $\mathbf x\in\mathbb R^p$ and a tensor $\mathcal X$, define the vector and tensor spatial-sign maps by
\[
\mathfrak{s}(\mathbf x)=
\begin{cases}
\mathbf x/\|\mathbf x\|_2,&\mathbf x\ne\mathbf0,\\
\mathbf0,&\mathbf x=\mathbf0,
\end{cases}
\]
\[
\mathfrak{s}_F(\mathcal X)=
\begin{cases}
\mathcal X/\|\mathcal X\|_F,&\mathcal X\ne\mathcal O,\\
\mathcal O,&\mathcal X=\mathcal O.
\end{cases}
\]
where $\mathcal O$ is the zero tensor of the required dimensions. These definitions satisfy
\[
\vecc\{\mathfrak{s}_F(\mathcal X)\}=\mathfrak{s}\{\vecc(\mathcal X)\}.
\]
A sample spatial median is any minimizer
\[
\widehat{\boldsymbol\mu}
\in\arg\min_{\boldsymbol\theta\in\mathbb R^p}
\sum_{m=1}^M\|\mathbf x_m-\boldsymbol\theta\|_2.
\]
Given a center $\boldsymbol\theta$, the spatial-sign covariance matrix is the average of $\mathfrak{s}(\mathbf x_m-\boldsymbol\theta)\mathfrak{s}(\mathbf x_m-\boldsymbol\theta)^T$. Each summand has spectral norm at most one, so the direct contribution of any single observation is bounded. Under ellipticity, the population spatial-sign covariance matrix and the shape matrix have the same eigenvectors and eigenvalue ordering, but they are not generally proportional \citep{durre2016eigenvalues}.

\subsection{Separable tensor elliptical distributions}
Let $p=\prod_{n=1}^N P_n$. We say that $\mathcal X\in\mathbb R^{P_1\times\cdots\times P_N}$ follows a separable tensor elliptical model if
\begin{equation}\label{eq:tensor-elliptical}
\begin{aligned}
\mathbf x=\vecc(\mathcal X)
&=\boldsymbol\mu+R\boldsymbol\Sigma^{1/2}\mathbf u,\\
\boldsymbol\Sigma
&=\boldsymbol\Sigma_N\otimes\cdots\otimes\boldsymbol\Sigma_1.
\end{aligned}
\end{equation}
where $\mathbf u$ is uniform on the unit sphere $\mathbb S^{p-1}$, $0<R<\infty$ almost surely, $R$ is independent of $\mathbf u$ and each $\boldsymbol\Sigma_n$ is positive definite. The Kronecker order agrees with \eqref{eq:vectorization-identity}. The mode scale factors are identifiable only up to reciprocal rescaling, and we impose $\tr(\boldsymbol\Sigma_n)=P_n$. The radius controls magnitude and tail thickness, whereas the mode scales determine the directional geometry. Whenever $R>0$, the spatial sign removes $R$ exactly.

\section{Spatial-sign-based multilinear principal component analysis}

\subsection{Working representation}
To describe the low-dimensional representation and specify the simulation design, consider the following low-rank-plus-noise model:
\begin{equation}\label{eq:working-model}
\mathcal X_m
=\mathcal C+\mathcal G_m\times_1\mathbf A_1\cdots\times_N\mathbf A_N+\mathcal E_m,
\end{equation}
where $\mathcal C$ is a center tensor, $\mathcal G_m\in\mathbb R^{K_1\times\cdots\times K_N}$ is a latent core tensor and $\mathbf A_n^T\mathbf A_n=\mathbf I_{K_n}$. Throughout, $1\le K_n<P_n$ for every mode $n$. The algorithm does not require a distributional specification for \eqref{eq:working-model}. The theoretical analysis instead assumes directly that the observed vectorized tensor satisfies \eqref{eq:tensor-elliptical}. This distinction matters because the sum of an elliptical low-rank signal and independent elliptical noise need not follow a separable elliptical distribution.

\subsection{Robust centering and sign transformation}
A tensor spatial median is any minimizer
\begin{equation}\label{eq:tensor-spatial-median}
\widehat{\mathcal C}
\in\arg\min_{\mathcal A\in\mathbb R^{P_1\times\cdots\times P_N}}
\sum_{m=1}^M\|\mathcal X_m-\mathcal A\|_F.
\end{equation}
Because vectorization preserves the Frobenius norm,
\[
\widehat{\boldsymbol\mu}:=\vecc(\widehat{\mathcal C})
\]
is a sample spatial median of $\{\vecc(\mathcal X_m):1\le m\le M\}$. Under Assumption~\ref{ass:main}(a), the observations are not contained in one affine line almost surely when $M\ge3$; the objective in \eqref{eq:tensor-spatial-median} is then strictly convex and the spatial median is unique. The argument is given in Appendix~B. For an arbitrary degenerate data set, a fixed deterministic rule may be used to select a minimizer; every deterministic inequality below holds for each minimizer separately.
For $\widetilde{\mathcal X}_m=\mathcal X_m-\widehat{\mathcal C}$, define
\begin{equation}\label{eq:tensor-sign}
\widehat{\mathcal S}_m=\mathfrak{s}_F(\widetilde{\mathcal X}_m).
\end{equation}
Thus $\|\widehat{\mathcal S}_m\|_F\le1$. Consequently, the direct contribution of each observation to the criterion is bounded; for every nonzero centered observation, the transformation preserves its direction in tensor space.

\subsection{Initialization and alternating optimization}
SMPCA estimates orthonormal loading matrices by
\begin{equation}\label{eq:smpca-objective}
\begin{aligned}
&\max_{\substack{\mathbf V_n^T\mathbf V_n=\mathbf I_{K_n}\\1\le n\le N}}
\widehat\Psi(\mathbf V_1,\ldots,\mathbf V_N)\\
&\qquad=\frac1M\sum_{m=1}^M
\|\widehat{\mathcal S}_m\times_1\mathbf V_1^T
\cdots\times_N\mathbf V_N^T\|_F^2.
\end{aligned}
\end{equation}
Let $\widehat{\mathbf s}_m=\vecc(\widehat{\mathcal S}_m)$ and $\mathbf P_n=\mathbf V_n\mathbf V_n^T$. The same criterion, regarded as a function of the loading projectors, is
\[
\widehat\Psi(\mathbf P_1,\ldots,\mathbf P_N)
=\frac1M\sum_{m=1}^M
\widehat{\mathbf s}_m^T
(\mathbf P_N\otimes\cdots\otimes\mathbf P_1)
\widehat{\mathbf s}_m.
\]
This representation makes explicit that the criterion is invariant to rotations of the columns of each loading matrix. Let $\widehat{\mathbf S}_{m(n)}$ be the mode-$n$ unfolding of $\widehat{\mathcal S}_m$ and define the initial mode scatter
\begin{equation}\label{eq:initial-mode-scatter}
\widehat{\boldsymbol\Gamma}_n^{\mathrm{init}}
=\frac1M\sum_{m=1}^M
\widehat{\mathbf S}_{m(n)}\widehat{\mathbf S}_{m(n)}^T.
\end{equation}
The initial loading matrix $\widehat{\mathbf V}_n^{(0)}$ contains the $K_n$ leading eigenvectors of $\widehat{\boldsymbol\Gamma}_n^{\mathrm{init}}$.

For a cyclic update of mode $n$, hold the other loading matrices fixed and form
\begin{equation}\label{eq:partial-projection}
\widehat{\mathcal Z}_{m,n}
=\widehat{\mathcal S}_m\mathop{\times}_{j\ne n}\mathbf V_j^T.
\end{equation}
Writing $\widehat{\mathbf Z}_{m,n}^{(n)}$ for its mode-$n$ unfolding, define
\begin{equation}\label{eq:block-scatter}
\begin{aligned}
\widehat{\boldsymbol\Gamma}_n^{\mathrm{blk}}(\mathsf P_{-n})
&=\frac1M\sum_{m=1}^M
\widehat{\mathbf Z}_{m,n}^{(n)}
\bigl(\widehat{\mathbf Z}_{m,n}^{(n)}\bigr)^T,\\
\mathsf P_{-n}
&=(\mathbf P_1,\ldots,\mathbf P_{n-1},
\mathbf P_{n+1},\ldots,\mathbf P_N).
\end{aligned}
\end{equation}
Although \eqref{eq:block-scatter} is computed from orthonormal bases $\mathbf V_j$, it depends on them only through the ordered projector tuple $\mathsf P_{-n}$. We write $\widehat\Psi(\mathbf P_n,\mathsf P_{-n})$ for the objective with these projectors held fixed. Then
\[
\widehat\Psi(\mathbf P_n,\mathsf P_{-n})
=\tr\{\mathbf P_n\widehat{\boldsymbol\Gamma}_n^{\mathrm{blk}}(\mathsf P_{-n})\}.
\]
Any rank-$K_n$ invariant subspace spanned by eigenvectors associated with the $K_n$ largest eigenvalues of
$\widehat{\boldsymbol\Gamma}_n^{\mathrm{blk}}(\mathsf P_{-n})$ is therefore a block maximizer. If the boundary eigenvalue is tied, a fixed deterministic rule selects one maximizing projector; any orthonormal basis of its range may be reported as the loading matrix. This convention makes a complete cyclic update a well-defined map, while all statistical bounds below remain valid for every maximizing selection.

\begin{algorithm*}[t]
\caption{Spatial-sign-based multilinear principal component analysis}
\label{alg:SMPCA}
\begin{algorithmic}[1]
\STATE \textbf{Input:} $\{\mathcal X_m\}_{m=1}^M$, dimensions $(K_1,\ldots,K_N)$, tolerance $\varepsilon_{\rm tol}>0$ and maximum number of sweeps $T_{\max}$.
\STATE \textbf{Output:} loading matrices $\{\widehat{\mathbf V}_n\}_{n=1}^N$ and core tensors $\{\widehat{\mathcal G}_m\}_{m=1}^M$.
\STATE Compute the spatial median in \eqref{eq:tensor-spatial-median} and the signs in \eqref{eq:tensor-sign}.
\STATE For each $n$, compute \eqref{eq:initial-mode-scatter} and set $\widehat{\mathbf V}_n^{(0)}$ to its $K_n$ leading eigenvectors.
\STATE Set $\widehat\Psi_0=\widehat\Psi(\widehat{\mathbf V}_1^{(0)},\ldots,\widehat{\mathbf V}_N^{(0)})$.
\FOR{$t=1,\ldots,T_{\max}$}
  \FOR{$n=1,\ldots,N$}
    \STATE Compute \eqref{eq:partial-projection} and \eqref{eq:block-scatter} using the most recently updated loading matrices.
    \STATE Replace $\widehat{\mathbf V}_n$ by the $K_n$ leading eigenvectors of $\widehat{\boldsymbol\Gamma}_n^{\mathrm{blk}}(\mathsf P_{-n})$.
  \ENDFOR
  \STATE Set $\widehat\Psi_t=\widehat\Psi(\widehat{\mathbf V}_1,\ldots,\widehat{\mathbf V}_N)$.
  \IF{$|\widehat\Psi_t-\widehat\Psi_{t-1}|<\varepsilon_{\rm tol}$}
     \STATE terminate.
  \ENDIF
\ENDFOR
\STATE Set $\widehat{\mathcal G}_m=(\mathcal X_m-\widehat{\mathcal C})\times_1\widehat{\mathbf V}_1^T\cdots\times_N\widehat{\mathbf V}_N^T$.
\end{algorithmic}
\end{algorithm*}

The normalization by $M$ leaves the eigenvectors unchanged and gives $0\le\widehat\Psi\le1$. Theorem~\ref{thm:algorithmic-convergence} establishes convergence of the objective values, whereas Theorem~\ref{thm:als-statistical} controls the statistical error of every exact block update. These are distinct properties: monotonicity of the objective does not by itself imply convergence of the loading matrices, a geometric rate or global optimality of a sample fixed point.

\subsection{Selection of the multilinear dimensions}
Let $\widehat\gamma_{n,1}\ge\cdots\ge\widehat\gamma_{n,P_n}\ge0$ be the eigenvalues of \eqref{eq:initial-mode-scatter}. When their sum is positive, define
\[
\widehat c_{n,0}=0,
\qquad
\widehat c_{n,k}
=\frac{\sum_{j=1}^k\widehat\gamma_{n,j}}
{\sum_{j=1}^{P_n}\widehat\gamma_{n,j}},
\qquad 1\le k\le P_n,
\]
and, for $\tau\in(0,1)$,
\begin{equation}\label{eq:rank-rule}
\widehat K_n^\circ(\tau)=\min\{k:\widehat c_{n,k}\ge\tau\}.
\end{equation}
This rule estimates the dimension defined by the corresponding population cumulative contribution. It may return $P_n$, in which case mode $n$ is retained without dimension reduction; the eigengap results for a non-trivial loading subspace are invoked only when the selected dimension is smaller than $P_n$. A fixed threshold, such as $0.95$, does not automatically recover an algebraic tensor rank; Theorem~\ref{thm:rank-consistency} gives the required separation condition.

\section{Theoretical properties}

\subsection{Targets, assumptions and stochastic rates}
Let
\[
\mathbf x_m=\vecc(\mathcal X_m),
\qquad
\boldsymbol\mu=\vecc(\mathcal C),
\qquad
p=\prod_{n=1}^N P_n,
\]
and retain $\widehat{\boldsymbol\mu}=\vecc(\widehat{\mathcal C})$ from Section~3.2. The tensor order $N$ is fixed, whereas $p$, the mode dimensions $P_n$ and the prescribed dimensions $K_n$ may depend on $M$ unless a fixed-dimensional regime is stated explicitly. Write
\[
\begin{aligned}
\boldsymbol\Sigma_n
&=\mathbf Q_n\diag(\lambda_{n,1},\ldots,\lambda_{n,P_n})\mathbf Q_n^T,\\
\lambda_{n,1}&\ge\cdots\ge\lambda_{n,P_n}>0.
\end{aligned}
\]
and let $\mathbf q_{n,j}$ be the $j$th column of $\mathbf Q_n$. For prescribed $1\le K_n<P_n$, define
\[
\begin{aligned}
\mathbf K&=(K_1,\ldots,K_N),\\
\mathbf V_n^\star&=[\mathbf q_{n,1},\ldots,\mathbf q_{n,K_n}],\\
\mathbf P_n^\star&=\mathbf V_n^\star\mathbf V_n^{\star T}.
\end{aligned}
\]
\[
K=\prod_{n=1}^N K_n,
\qquad
K_{-n}=\prod_{j\ne n}K_j.
\]
For two loading matrices with the same number of columns, write
\[
d(\mathbf V,\mathbf W)
=\frac1{\sqrt2}
\|\mathbf V\mathbf V^T-\mathbf W\mathbf W^T\|_F.
\]
The rank-$K_n$ projector space is
\[
\begin{aligned}
\mathfrak P_{n,K_n}
&=\{\mathbf P=\mathbf P^T=\mathbf P^2:
\tr(\mathbf P)=K_n\},\\
\mathfrak P_{\mathbf K}
&=\prod_{n=1}^N\mathfrak P_{n,K_n}.
\end{aligned}
\]
For an ordered tuple $\mathsf P=(\mathbf P_1,\ldots,\mathbf P_N)\in\mathfrak P_{\mathbf K}$, let $\mathsf P_{-n}$ be the tuple with its $n$th component removed and let $\mathfrak P_{-n,\mathbf K}$ be the corresponding product space. Empty products are one. The target joint projector is
\[
\boldsymbol\Pi^\star
=\mathbf P_N^\star\otimes\cdots\otimes\mathbf P_1^\star.
\]

Let
\[
\begin{aligned}
\mathcal I&=\prod_{n=1}^N\{1,\ldots,P_n\},\\
\mathbf Q&=\mathbf Q_N\otimes\cdots\otimes\mathbf Q_1,\\
\nu_{\mathbf i}&=\prod_{n=1}^N\lambda_{n,i_n}.
\end{aligned}
\]
where $\mathbf i=(i_1,\ldots,i_N)$. For independent standard normal variables $\{Z_{\mathbf i}:\mathbf i\in\mathcal I\}$, define
\begin{equation}\label{eq:omega-index}
\omega_{\mathbf i}
=\E\left(
\frac{\nu_{\mathbf i}Z_{\mathbf i}^2}
{\sum_{\mathbf j\in\mathcal I}\nu_{\mathbf j}Z_{\mathbf j}^2}
\right).
\end{equation}
For mode $n$, $\mathbf i_{-n}$ is the ordered multi-index obtained by deleting $i_n$, and $(a,\mathbf i_{-n})$ inserts $a$ in the $n$th position. Put
\[
\begin{aligned}
\gamma_{n,a}
&=\sum_{\mathbf i_{-n}}\omega_{(a,\mathbf i_{-n})},\\
\Delta_n^{\Sigma}
&=\lambda_{n,K_n}-\lambda_{n,K_n+1},\\
\Delta_n^{\mathrm{sgn}}
&=\min_{\mathbf i_{-n}}
\{\omega_{(K_n,\mathbf i_{-n})}
-\omega_{(K_n+1,\mathbf i_{-n})}\},\\
\Delta_n^{\mathrm{init}}
&=\gamma_{n,K_n}-\gamma_{n,K_n+1}.
\end{aligned}
\]
and $\Delta_{\min}^{\mathrm{sgn}}=\min_n\Delta_n^{\mathrm{sgn}}$.

For $\mathbf s_m=\mathfrak{s}(\mathbf x_m-\boldsymbol\mu)$, let $\mathcal S_m=\unvec(\mathbf s_m)$ and let $\mathbf S_{m(n)}$ be its mode-$n$ unfolding. Define
\[
\begin{aligned}
\boldsymbol\Omega
&=\E(\mathbf s_m\mathbf s_m^T),\\
\boldsymbol\Gamma_n^{\mathrm{init}}
&=\E\{\mathbf S_{m(n)}\mathbf S_{m(n)}^T\},\\
\widehat{\boldsymbol\Omega}
&=\frac1M\sum_{m=1}^M
\widehat{\mathbf s}_m\widehat{\mathbf s}_m^T.
\end{aligned}
\]
The population criterion is
\[
\Psi(\mathsf P)
=\tr\{\boldsymbol\Omega(\mathbf P_N\otimes\cdots\otimes\mathbf P_1)\}.
\]

\begin{assumption}\label{ass:main}
\begin{enumerate}[label=(\alph*)]
\item The observations are independent and identically distributed, with
\[
\mathbf x_m=\boldsymbol\mu+R_m\boldsymbol\Sigma^{1/2}\mathbf u_m,
\qquad
\boldsymbol\Sigma=\boldsymbol\Sigma_N\otimes\cdots\otimes\boldsymbol\Sigma_1,
\]
where $R_m>0$ is independent of $\mathbf u_m\sim\operatorname{Unif}(\mathbb S^{p-1})$, $p\ge2$, and $\boldsymbol\Sigma_n$ is positive definite with $\tr(\boldsymbol\Sigma_n)=P_n$.
\item Writing $R_m=\sqrt p\,\xi_m$, there are constants $c_\xi,C_\xi>0$, independent of $M$ and $p$, such that
\[
\xi_m>0\ \text{a.s.},
\qquad
c_\xi\le\E(\xi_m^{-1}),
\qquad
\E(\xi_m^{-2})\le C_\xi.
\]
\item $\kappa(\boldsymbol\Sigma)\le\kappa_0$ for a constant $\kappa_0\ge1$ independent of $M$ and $p$.
\item $\Delta_n^{\Sigma}>0$ for $1\le n\le N$.
\end{enumerate}
\end{assumption}

Assumption~\ref{ass:main}(a) specifies a separable elliptical model. Under Assumption~\ref{ass:main}(a), Theorem~\ref{thm:population-structure} shows that \(\boldsymbol\Omega\) has eigenvectors \(\mathbf Q\) and eigenvalues \(\omega_{\mathbf i}\), while \(\boldsymbol\Gamma_n^{\mathrm{init}}\) has eigenvectors \(\mathbf Q_n\) and eigenvalues \(\gamma_{n,1},\ldots,\gamma_{n,P_n}\). The eigenvalues \(\omega_{\mathbf i}\) are independent of the radial distribution and preserve both the strict ordering and the ties of the product shape eigenvalues. Part (b) controls the inverse radial moments needed for spatial-median and feasible spatial-sign covariance estimation while imposing no positive-moment condition on $R_m$. Part (c) rules out increasingly ill-conditioned shape matrices, and part (d) identifies the leading loading space in each mode. 

Appendix~\ref{sec:covered-distributions} verifies Assumption~\ref{ass:main} for separable tensor $t$ distributions and for a common-center, proportional-scale subclass of tensor-normal mixtures. The common-center and proportional-scale restrictions cannot be dropped in general because an arbitrary Gaussian mixture need not be elliptically contoured.

For $q\ge1$, write
\[
\chi_M(q)
=\sqrt{\frac{\log(2q)}{M}}
+\frac{\log(2q)}{M}
+\frac1{\sqrt M}.
\]
Appendix~\ref{sec:auxiliary} proves that, under Assumptions~\ref{ass:main}(a)--(c) and $p/M\to0$,
\[
\begin{aligned}
\|\widehat{\boldsymbol\mu}-\boldsymbol\mu\|_2
&=O_P\left(\sqrt{\frac pM}\right),\\
\|\widehat{\boldsymbol\Gamma}_n^{\mathrm{init}}
-\boldsymbol\Gamma_n^{\mathrm{init}}\|_{\op}
&=O_P\{\chi_M(P_n)\},\\
\|\widehat{\boldsymbol\Omega}-\boldsymbol\Omega\|_{\op}
&=O_P\{\chi_M(p)\}.
\end{aligned}
\]

\subsection{Tensor-specific population geometry}
The following results use the product eigenbasis of the separable tensor model rather than treating $\vecc(\mathcal X_m)$ as an unrestricted $p$-dimensional vector.

\begin{proposition}[Inheritance of mode-wise eigengaps]\label{prop:sign-gap}
Under Assumptions~\ref{ass:main}(a),(d), for every mode $n$ the positive shape eigengap is inherited by both relevant spatial-sign spectra:
\[
\Delta_n^{\mathrm{sgn}}>0,
\qquad
\Delta_n^{\mathrm{init}}>0.
\]
If Assumption~\ref{ass:main}(c) also holds, then
\begin{equation}\label{eq:gap-lower-bounds}
\begin{aligned}
\Delta_n^{\mathrm{sgn}}
&\ge\frac{\Delta_n^{\Sigma}}{\kappa_0^5(p+2)},\\
\Delta_n^{\mathrm{init}}
&\ge\frac p{P_n}\Delta_n^{\mathrm{sgn}}
\ge\frac p{p+2}
\frac{\Delta_n^{\Sigma}}{\kappa_0^5P_n}.
\end{aligned}
\end{equation}
Thus no separate population eigengap assumption is required for the spatial-sign covariance matrix or the initial mode scatter.
\end{proposition}

For $\mathsf P_{-n}\in\mathfrak P_{-n,\mathbf K}$ and
$\mathbf v\in\mathbb R^{P_n}$, define
\[
\begin{aligned}
\mathbf B_n(\mathbf v;\mathsf P_{-n})
&=\mathbf P_N\otimes\cdots\otimes\mathbf P_{n+1}
\otimes\mathbf v\mathbf v^T\\
&\quad\otimes\mathbf P_{n-1}\otimes\cdots\otimes\mathbf P_1.
\end{aligned}
\]
The population block scatter is the unique symmetric matrix satisfying
\[
\mathbf v^T\boldsymbol\Gamma_n^{\mathrm{blk}}(\mathsf P_{-n})\mathbf v
=\tr\{\boldsymbol\Omega\mathbf B_n(\mathbf v;\mathsf P_{-n})\},
\qquad \mathbf v\in\mathbb R^{P_n}.
\]
Let
\[
\begin{aligned}
\Delta_n^{\mathrm{blk}}(\mathsf P_{-n})
&=\lambda_{K_n}
\{\boldsymbol\Gamma_n^{\mathrm{blk}}(\mathsf P_{-n})\}\\
&\quad-\lambda_{K_n+1}
\{\boldsymbol\Gamma_n^{\mathrm{blk}}(\mathsf P_{-n})\}.
\end{aligned}
\]
Let
\[
\begin{aligned}
w_{j,a}
&=\mathbf q_{j,a}^T\mathbf P_j\mathbf q_{j,a},\\
g_{n,a}(\mathsf P_{-n})
&=\sum_{\mathbf i_{-n}}\omega_{(a,\mathbf i_{-n})}
\prod_{j\ne n}w_{j,i_j}.
\end{aligned}
\]

\begin{theorem}[Tensor contraction and one-sweep population recovery]\label{thm:tensor-contraction}
Under Assumptions~\ref{ass:main}(a),(d), fix a mode $n$ and any ordered tuple $\mathsf P_{-n}\in\mathfrak P_{-n,\mathbf K}$. Then
\begin{equation}\label{eq:tensor-contraction-spectrum}
\boldsymbol\Gamma_n^{\mathrm{blk}}(\mathsf P_{-n})
=\mathbf Q_n
\diag\{g_{n,a}(\mathsf P_{-n})\}_{a=1}^{P_n}
\mathbf Q_n^T.
\end{equation}
The contracted eigenvalues preserve the ordering, including ties, of the mode-$n$ shape eigenvalues. In particular,
\begin{equation}\label{eq:uniform-block-gap}
\Delta_n^{\mathrm{blk}}(\mathsf P_{-n})
\ge K_{-n}\Delta_n^{\mathrm{sgn}},
\end{equation}
and every rank-$K_n$ projector $\mathbf P_n=\mathbf V_n\mathbf V_n^T$ satisfies
\begin{equation}\label{eq:block-curvature}
\Psi(\mathbf P_n^\star,\mathsf P_{-n})-\Psi(\mathbf P_n,\mathsf P_{-n})
\ge K_{-n}\Delta_n^{\mathrm{sgn}}d(\mathbf V_n,\mathbf V_n^\star)^2.
\end{equation}
Consequently, the population mode-$n$ update has the unique solution $\mathbf P_n^\star$ for every choice of the other current projectors. Hence one complete cyclic population sweep recovers the target tuple from any initial tuple.
\end{theorem}

The representation in \eqref{eq:tensor-contraction-spectrum} is a weighted marginalization of the product-basis eigenvalues of the spatial-sign covariance matrix, not a Kronecker factorization of $\boldsymbol\Omega$. The one-sweep conclusion is a population property and does not assert exact one-sweep recovery by the finite-sample algorithm.

\begin{corollary}[Global identification and multilinear curvature]\label{cor:population-curvature}
Under the conditions of Theorem~\ref{thm:tensor-contraction}, every tuple $\mathsf P=(\mathbf P_1,\ldots,\mathbf P_N)\in\mathfrak P_{\mathbf K}$, with $\mathbf P_n=\mathbf V_n\mathbf V_n^T$, satisfies
\begin{equation}\label{eq:global-curvature}
\Psi(\mathbf P_1^\star,\ldots,\mathbf P_N^\star)-\Psi(\mathsf P)
\ge\sum_{n=1}^N K_{-n}\Delta_n^{\mathrm{sgn}}d(\mathbf V_n,\mathbf V_n^\star)^2.
\end{equation}
The right-hand side vanishes only when $\mathbf P_n=\mathbf P_n^\star$ for all $n$. Therefore the target tuple is the unique global maximizer of the population SMPCA criterion.
\end{corollary}

\subsection{SMPCA versus vectorized PCA and spatial-sign PCA}\label{sec:target-comparison}
Vectorized spatial-sign PCA maximizes $\tr(\mathbf U^T\boldsymbol\Omega\mathbf U)$ over $\mathbf U^T\mathbf U=\mathbf I_K$, whereas SMPCA restricts the corresponding rank-$K$ projector to the Kronecker-structured form $\mathbf P_N\otimes\cdots\otimes\mathbf P_1$. Define
\[
\mathcal I_{\mathbf K}=\prod_{n=1}^N\{1,\ldots,K_n\},
\qquad
\nu_{\mathrm{in}}(\mathbf K)=\prod_{n=1}^N\lambda_{n,K_n},
\]
\[
\nu_{\mathrm{out}}(\mathbf K)
=\max_{1\le n_0\le N}
\left\{\lambda_{n_0,K_{n_0}+1}\prod_{j\ne n_0}\lambda_{j,1}\right\}.
\]

\begin{proposition}[Multilinear versus unrestricted vectorized targets]\label{prop:rectangular-target}
Under Assumption~\ref{ass:main}(a), compare the rectangular multilinear space $\operatorname{range}(\boldsymbol\Pi^\star)$ with the unrestricted leading rank-$K$ eigenspace of both $\boldsymbol\Sigma$ and $\boldsymbol\Omega$. It is the unique leading eigenspace when $\nu_{\mathrm{in}}(\mathbf K)>\nu_{\mathrm{out}}(\mathbf K)$, one of several leading eigenspaces when equality holds, and different from every leading eigenspace when $\nu_{\mathrm{in}}(\mathbf K)<\nu_{\mathrm{out}}(\mathbf K)$.
If Assumption~\ref{ass:main}(d) also holds, each mode loading space remains uniquely identifiable by SMPCA in all three cases. Thus mode-wise identifiability does not require the multilinear target to coincide with the unrestricted vectorized target.
\end{proposition}

If $\E(R_m^2)<\infty$, $\operatorname{Cov}(\mathbf x_m)=\E(R_m^2)\boldsymbol\Sigma/p$, so the same trichotomy applies to ordinary vectorized PCA. The multilinear parameter has structural dimension
\[
\mathrm{df}_{\mathrm{ten}}=\sum_{n=1}^N K_n(P_n-K_n),
\]
whereas an unrestricted rank-$K$ subspace has dimension $\mathrm{df}_{\mathrm{vec}}=K(p-K)$. For fixed $K_1,\ldots,K_N$, $N\ge2$ and $\min_nP_n\to\infty$, $\mathrm{df}_{\mathrm{ten}}/\mathrm{df}_{\mathrm{vec}}\to0$.

\subsection{Statistical properties}
For $\mathbf P_n=\mathbf V_n\mathbf V_n^T$, put
\[
\begin{aligned}
\boldsymbol\Pi
&=\mathbf P_N\otimes\cdots\otimes\mathbf P_1,\\
d_n&=d(\mathbf V_n,\mathbf V_n^\star),\\
d_\otimes(\boldsymbol\Pi,\boldsymbol\Pi^\star)
&=\frac{\|\boldsymbol\Pi-\boldsymbol\Pi^\star\|_F}{\sqrt{2K}}.
\end{aligned}
\]
The exact identity
\[
d_\otimes(\boldsymbol\Pi,\boldsymbol\Pi^\star)^2
=1-\prod_{n=1}^N\left(1-\frac{d_n^2}{K_n}\right)
\]
and its deterministic consequences are collected in Proposition~\ref{prop:projector-geometry}.

For every estimated loading matrix, write
$\widehat{\mathbf P}_n=\widehat{\mathbf V}_n\widehat{\mathbf V}_n^T$ and
$\widehat{\boldsymbol\Pi}=\widehat{\mathbf P}_N\otimes\cdots\otimes\widehat{\mathbf P}_1$. In particular, let
\[
\begin{aligned}
\varepsilon_{n,M}^{\mathrm{init}}
&=\|\widehat{\boldsymbol\Gamma}_n^{\mathrm{init}}
-\boldsymbol\Gamma_n^{\mathrm{init}}\|_{\op},\\
\widehat{\boldsymbol\Pi}^{(0)}
&=\widehat{\mathbf P}_N^{(0)}\otimes\cdots
\otimes\widehat{\mathbf P}_1^{(0)}.
\end{aligned}
\]

\begin{theorem}[Initial loading-space error]\label{thm:initial-subspace}
Under Assumptions~\ref{ass:main}(a),(d), let $\widehat{\mathbf V}_n^{(0)}$ be an orthonormal basis of any leading rank-$K_n$ eigenspace of $\widehat{\boldsymbol\Gamma}_n^{\mathrm{init}}$. Then, for every mode $n$,
\begin{equation}\label{eq:initial-mode-bound}
d(\widehat{\mathbf V}_n^{(0)},\mathbf V_n^\star)
\le\frac{2\sqrt{K_n}}{\Delta_n^{\mathrm{init}}}\varepsilon_{n,M}^{\mathrm{init}},
\end{equation}
and the resulting joint projector obeys
\begin{equation}\label{eq:initial-joint-bound}
d_\otimes(\widehat{\boldsymbol\Pi}^{(0)},\boldsymbol\Pi^\star)
\le2\left\{\sum_{n=1}^N
\left(\frac{\varepsilon_{n,M}^{\mathrm{init}}}{\Delta_n^{\mathrm{init}}}\right)^2\right\}^{1/2}.
\end{equation}
These inequalities hold for any leading sample eigenspace; Proposition~\ref{prop:empirical-gaps} gives a sufficient condition for its uniqueness. If Assumptions~\ref{ass:main}(b),(c) also hold and $p/M\to0$, the two bounds are, respectively,
\[
O_P\left\{
\frac{\sqrt{K_n}\chi_M(P_n)}{\Delta_n^{\mathrm{init}}}
\right\}
\]
and
\[
O_P\left[\left\{\sum_{n=1}^N
\frac{\chi_M(P_n)^2}
{(\Delta_n^{\mathrm{init}})^2}\right\}^{1/2}\right].
\]
\end{theorem}

Let $\widehat{\mathbf V}_n^{\mathrm{new}}(\mathsf P_{-n})$ denote any orthonormal basis returned by an exact sample block update at $\mathsf P_{-n}$.

\begin{theorem}[Uniform error of exact block updates]\label{thm:als-statistical}
Under Assumptions~\ref{ass:main}(a),(d), fix a mode $n$. For every current tuple $\mathsf P_{-n}\in\mathfrak P_{-n,\mathbf K}$, any leading rank-$K_n$ eigenspace returned by the exact sample block update satisfies
\begin{equation}\label{eq:als-update-bound}
d\{\widehat{\mathbf V}_n^{\mathrm{new}}(\mathsf P_{-n}),\mathbf V_n^\star\}
\le\frac{2\sqrt{K_n}}{\Delta_n^{\mathrm{sgn}}}
\|\widehat{\boldsymbol\Omega}-\boldsymbol\Omega\|_{\op}.
\end{equation}
The bound is uniform over the current projectors and therefore applies to the data-dependent iterates of Algorithm~\ref{alg:SMPCA}. After any complete sweep, in which every mode has been updated once,
\begin{equation}\label{eq:als-joint-bound}
d_\otimes(\widehat{\boldsymbol\Pi},\boldsymbol\Pi^\star)
\le2\|\widehat{\boldsymbol\Omega}-\boldsymbol\Omega\|_{\op}
\left\{\sum_{n=1}^N(\Delta_n^{\mathrm{sgn}})^{-2}\right\}^{1/2}.
\end{equation}
Proposition~\ref{prop:empirical-gaps} gives a uniform event on which all sample block projectors are unique. Under Assumptions~\ref{ass:main}(b),(c) and $p/M\to0$, the mode-wise and joint bounds are, respectively,
\[
O_P\left\{
\frac{\sqrt{K_n}\chi_M(p)}{\Delta_n^{\mathrm{sgn}}}
\right\}
\]
and
\[
O_P\left[\chi_M(p)
\left\{\sum_{n=1}^N
(\Delta_n^{\mathrm{sgn}})^{-2}\right\}^{1/2}\right].
\]
\end{theorem}

Empirical eigengap bounds, uniqueness of the block solutions and the same error bounds for coordinatewise and global sample maximizers are stated in Propositions~\ref{prop:empirical-gaps} and~\ref{prop:sample-solutions}.

\begin{corollary}[Consistency of loading spaces and the joint projector]\label{cor:loading-consistency}
Suppose Assumptions~\ref{ass:main}(a)--(d) hold, $p/M\to0$ and $N$ is fixed. If
\begin{equation}\label{eq:initial-consistency-condition}
\sum_{n=1}^N\frac{\chi_M(P_n)^2}{(\Delta_n^{\mathrm{init}})^2}\longrightarrow0,
\end{equation}
then the initial joint projector is consistent in the rank-normalized distance $d_\otimes$:
\[
d_\otimes(\widehat{\boldsymbol\Pi}^{(0)},\boldsymbol\Pi^\star)\xrightarrow{P}0.
\]
When the ranks grow, this normalized conclusion is distinct from mode-wise consistency or convergence in the unnormalized Frobenius norm; those follow only under the corresponding individual rates stated below. If
\begin{equation}\label{eq:als-consistency-condition}
\chi_M(p)\left\{\sum_{n=1}^N(\Delta_n^{\mathrm{sgn}})^{-2}\right\}^{1/2}\longrightarrow0,
\end{equation}
then the joint projector obtained after any complete exact sweep is consistent. The same conclusion holds for every coordinatewise maximizer and every global maximizer of the sample criterion. Mode-wise consistency follows whenever the corresponding individual rate in Theorem~\ref{thm:initial-subspace} or~\ref{thm:als-statistical} tends to zero.
\end{corollary}

Define
\[
\widehat{\mathcal R}(\mathbf x)
=\widehat{\boldsymbol\mu}+\widehat{\boldsymbol\Pi}(\mathbf x-\widehat{\boldsymbol\mu}),
\qquad
\mathcal R^\star(\mathbf x)
=\boldsymbol\mu+\boldsymbol\Pi^\star(\mathbf x-\boldsymbol\mu).
\]

\begin{corollary}[Joint projection and reconstruction]\label{cor:joint-projector}
Let $\widehat{\boldsymbol\Pi}$ be the joint projector obtained after a complete exact sweep. Under the conditions of Theorem~\ref{thm:als-statistical},
\begin{equation}\label{eq:joint-projector-bound}
\|\widehat{\boldsymbol\Pi}-\boldsymbol\Pi^\star\|_{\op}
\le2\sqrt2\,\|\widehat{\boldsymbol\Omega}-\boldsymbol\Omega\|_{\op}
\sum_{n=1}^N\frac{\sqrt{K_n}}{\Delta_n^{\mathrm{sgn}}}.
\end{equation}
If Assumptions~\ref{ass:main}(b),(c) also hold, $p/M\to0$, and $\mathbf x$ is deterministic, then
\[
\begin{aligned}
&\|\widehat{\mathcal R}(\mathbf x)
-\mathcal R^\star(\mathbf x)\|_2\\
&\quad=O_P\left[
\chi_M(p)\left\{\sum_{n=1}^N
\frac{\sqrt{K_n}}{\Delta_n^{\mathrm{sgn}}}\right\}
\|\mathbf x-\boldsymbol\mu\|_2
+\sqrt{\frac pM}\right].
\end{aligned}
\]
Thus reconstruction is consistent whenever the deterministic rate on the right tends to zero. Proposition~\ref{prop:reconstruction-details} covers sample-dependent inputs.
\end{corollary}

The pathwise deterministic inequality and the extension to data-dependent inputs are given in Proposition~\ref{prop:reconstruction-details}. Sufficient growth conditions expressed through $\Delta_n^\Sigma$ follow by substituting \eqref{eq:gap-lower-bounds}.

\subsection{Algorithmic convergence and dimension selection}

\begin{theorem}[Convergence of the objective values]\label{thm:algorithmic-convergence}
Consider the ideal exact cyclic sequence obtained from Algorithm~\ref{alg:SMPCA} by continuing the block updates without the tolerance or maximum-iteration stopping rules. If $\widehat\Psi_t$ is the sample criterion after the $t$th complete sweep, then
\[
0\le\widehat\Psi_0\le\widehat\Psi_1\le\cdots\le1,
\qquad
\widehat\Psi_t\longrightarrow\widehat\Psi_\infty\in[0,1].
\]
Every fixed point of a specified complete cyclic update map is a coordinatewise maximizer of \eqref{eq:smpca-objective}.
\end{theorem}

Note that the result asserts convergence of the objective values, not convergence of the projector iterates or global optimality of a fixed point. Under \eqref{eq:als-consistency-condition}, every fixed point satisfies the statistical bounds in Proposition~\ref{prop:sample-solutions}; no geometric convergence rate is claimed.

Let
\[
\begin{aligned}
c_{n,k}&=\sum_{j=1}^k\gamma_{n,j},\\
K_n^\circ(\tau)&=\min\{k:c_{n,k}\ge\tau\},\\
c_{n,0}&=0.
\end{aligned}
\]
and define the population threshold margin
\begin{equation}\label{eq:rank-margin}
\mathfrak m_n(\tau)
=\min\{\tau-c_{n,K_n^\circ(\tau)-1},\ c_{n,K_n^\circ(\tau)}-\tau\}.
\end{equation}

\begin{theorem}[Consistency of cumulative-contribution dimension selection]\label{thm:rank-consistency}
For the sample selector in \eqref{eq:rank-rule}, fix a mode $n$, assume $M\ge2$, and let $K_n^\circ(\tau)$ be the population dimension selected by the cumulative-contribution threshold $\tau$. Suppose the threshold is separated from the adjacent population cumulative contributions, so that $\mathfrak m_n(\tau)>0$. Under Assumptions~\ref{ass:main}(a)--(c) and $p/M\to0$,
\[
\Prob\{\widehat K_n^\circ(\tau)=K_n^\circ(\tau)\}\longrightarrow1
\]
whenever
\begin{equation}\label{eq:rank-growth-condition}
\frac{P_n\chi_M(P_n)}{\mathfrak m_n(\tau)}
+\frac1{M\mathfrak m_n(\tau)}\longrightarrow0.
\end{equation}
For fixed dimensions, it is sufficient that $\mathfrak m_n(\tau)$ be bounded away from zero. The target $K_n^\circ(\tau)$ is threshold-defined and need not equal an algebraic tensor rank.
\end{theorem}

\section{Simulation}
We compare SMPCA with MPCA \citep{lu2008mpca}, casewise and cellwise robust MPCA (ROMPCA) \citep{hirari2026robust}, and the TPCA-\(L_p\) procedures \citep{7924765}. The experiments examine loading-space estimation under light- and heavy-tailed generators, computation time for the reported implementations, and cumulative-contribution dimension selection under mixture contamination.

\subsection{Design}
We generate independent third-order tensors $\mathcal X_m\in\mathbb R^{P_1\times P_2\times P_3}$ from
\begin{equation}\label{eq:simulation-model}
\mathcal X_m
=\mathcal G_m\times_1\mathbf A_1\times_2\mathbf A_2\times_3\mathbf A_3
+\sigma_e\mathcal E_m.
\end{equation}
The dimensions are $(P_1,P_2,P_3)=(30,20,5)$ and the data-generating multilinear ranks are $(K_1,K_2,K_3)=(8,6,2)$. For mode $n$, the columns of $\mathbf A_n$ are the leading $K_n$ eigenvectors of the AR(1) matrix with entries $(-0.9)^{|i-j|}$. The $(i,j,k)$ entry of the initially generated core is multiplied by
$\{K_1K_2K_3/(ijk)\}^f$, with $f\in\{1/2,1/4,1/8\}$, and $\sigma_e=0.1$.

The core and noise entries are generated independently from one of the following three symmetric distributions, with the same distribution used for both:
\begin{enumerate}[label=(\alph*)]
\item standard Gaussian;
\item Student $t$ with $2.5$ degrees of freedom;
\item a centered Gaussian mixture with contamination probability $0.2$, baseline variance $0.1$ and variance-inflation factor $9$.
\end{enumerate}
These entrywise generators are symmetric. Except in the Gaussian case, however, independent coordinates from the stated univariate laws do not generally produce jointly elliptical core or noise vectorizations. Accordingly, the sum in \eqref{eq:simulation-model} need not follow a separable tensor elliptical distribution. The simulation design is therefore broader than Assumption~\ref{ass:main}; the theory applies directly when the distribution of the observed vectorized tensor satisfies \eqref{eq:tensor-elliptical}. The common-center proportional tensor-normal mixture treated in Appendix~\ref{sec:covered-distributions} uses one scalar scale for an entire tensor and is therefore different from the entrywise mixture used here.

We take $M\in\{100,200\}$. Because the simulated mode eigenvalues are ordered and separated, the individual directions are identifiable up to sign. Let
$\mathbf a_{nj}$ be the $j$th column of $\mathbf A_n$ and let
$\widehat{\mathbf a}_{nj}$ be the correspondingly ordered estimated loading direction. We report
\[
\operatorname{SEE}
=\frac{1}{\sum_{n=1}^3K_n}
\sum_{n=1}^3\sum_{j=1}^{K_n}
\{1-|\widehat{\mathbf a}_{nj}^T\mathbf a_{nj}|\}.
\]
For repeated or nearly repeated eigenvalues, the mode-$n$ projector distance $d(\widehat{\mathbf V}_n,\mathbf A_n)$, $1\le n\le3$, is the appropriate rotation-invariant criterion. We first use 100 Monte Carlo replications for all methods. Because ROMPCA and TPCA-\(L_p\) are much more expensive in the reported implementation, the 1000-replication comparison is restricted to MPCA and SMPCA.

\begin{table*}[t]
\centering
\small
\setlength{\tabcolsep}{3pt}
\caption{Mean subspace estimation error over 100 replications}
\label{tab:accuracy}
\begin{tabular}{@{}ccccc ccc ccc@{}}
\toprule
& & \multicolumn{3}{c}{Gaussian} & \multicolumn{3}{c}{$t_{2.5}$} & \multicolumn{3}{c}{Gaussian mixture} \\
\cmidrule(lr){3-5} \cmidrule(lr){6-8} \cmidrule(lr){9-11}
& $f$ & $0.5$ & $0.25$ & $0.125$ & $0.5$ & $0.25$ & $0.125$ & $0.5$ & $0.25$ & $0.125$ \\
\midrule

\multirow{6}{*}{$M=100$}
& MPCA      & 0.0346 & \textbf{0.0921} & \textbf{0.2208} & 0.1640 & 0.2915 & 0.4392 & 0.0816 & 0.1819 & 0.3433 \\
& SMPCA     & 0.0330 & 0.0939 & 0.2216 & \textbf{0.0321} & \textbf{0.1022} & \textbf{0.2440} & \textbf{0.0313} & \textbf{0.0913} & \textbf{0.2342} \\
& ROMPCA    & 0.0390 & 0.0948 & 0.2277 & 0.0416 & 0.1170 & 0.2708 & 0.0470 & 0.1283 & 0.2807 \\
& TPCA-\(L_{0.5}\) & 0.0837 & 0.2015 & 0.3561 & 0.0969 & 0.2237 & 0.3784 & 0.0866 & 0.2228 & 0.3520 \\
& TPCA-\(L_1\)   & 0.0590 & 0.1546 & 0.2980 & 0.0934 & 0.2233 & 0.3435 & 0.0701 & 0.1811 & 0.3303 \\
& TPCA-\(L_{1.5}\) & \textbf{0.0312} & 0.1032 & 0.2362 & 0.0987 & 0.2139 & 0.3742 & 0.0508 & 0.1542 & 0.3081 \\
\addlinespace
\multirow{6}{*}{$M=200$}
& MPCA      & 0.0137 & \textbf{0.0426} & \textbf{0.1421} & 0.1337 & 0.2602 & 0.3999 & 0.0381 & 0.1018 & 0.2446 \\
& SMPCA     & \textbf{0.0136} & 0.0437 & 0.1443 & \textbf{0.0143} & \textbf{0.0489} & \textbf{0.1578} & \textbf{0.0159} & \textbf{0.0479} & \textbf{0.1548} \\
& ROMPCA    & 0.0180 & 0.0520 & 0.1493 & 0.0176 & 0.0580 & 0.1771 & 0.0245 & 0.0713 & 0.2040 \\
& TPCA-\(L_{0.5}\) & 0.0503 & 0.1478 & 0.2836 & 0.0547 & 0.1658 & 0.3100 & 0.0524 & 0.1440 & 0.3040 \\
& TPCA-\(L_1\)   & 0.0286 & 0.0896 & 0.2306 & 0.0498 & 0.1421 & 0.2819 & 0.0422 & 0.1137 & 0.2606 \\
& TPCA-\(L_{1.5}\) & 0.0145 & 0.0568 & 0.1731 & 0.0674 & 0.1605 & 0.3196 & 0.0272 & 0.0844 & 0.2340 \\
\bottomrule
\end{tabular}
\end{table*}

\subsection{Loading-space accuracy and computation time}
Table~\ref{tab:accuracy} reports the mean SEE over 100 replications. Under Gaussian sampling, MPCA and SMPCA are very close. Across the \(t_{2.5}\) and Gaussian mixture settings, the error of MPCA increases markedly, whereas SMPCA remains stable. The difference is especially pronounced for $M=200$, for which the sign normalization prevents a small number of large tensor norms from dominating the mode scatters.

Across the heavy-tailed settings in Table~\ref{tab:accuracy}, SMPCA has smaller SEE than ROMPCA and each TPCA-\(L_p\) variant. Table~\ref{tab:1000reps} confirms, with 1000 replications, that the small Gaussian difference and the large heavy-tail advantage are not artifacts of the shorter experiment.

\begin{table*}[t]
\centering
\small 
\setlength{\tabcolsep}{3pt} 
\caption{Mean subspace estimation error over 1000 replications}
\label{tab:1000reps}
\begin{tabular}{@{}ccccc ccc ccc@{}}
\toprule
& & \multicolumn{3}{c}{Gaussian} & \multicolumn{3}{c}{$t_{2.5}$} & \multicolumn{3}{c}{Gaussian mixture} \\ 
\cmidrule(lr){3-5} \cmidrule(lr){6-8} \cmidrule(lr){9-11}
& $f$ & $1/2$ & $1/4$ & $1/8$ & $1/2$ & $1/4$ & $1/8$ & $1/2$ & $1/4$ & $1/8$ \\
\midrule
\multirow{2}{*}{$M=100$} 
& MPCA & \textbf{0.0329} & \textbf{0.0935} & \textbf{0.2281} & 0.1761 & 0.2980 & 0.4396 & 0.0832 & 0.1848 & 0.3408 \\
& SMPCA & 0.0335 & 0.0946 & 0.2288 & \textbf{0.0337} & \textbf{0.0974} & \textbf{0.2368} & \textbf{0.0326} & \textbf{0.0935} & \textbf{0.2356} \\
\addlinespace
\multirow{2}{*}{$M=200$} 
& MPCA & \textbf{0.0138} & \textbf{0.0456} & \textbf{0.1482} & 0.1413 & 0.2618 & 0.4048 & 0.0403 & 0.1088 & 0.2526 \\
& SMPCA & 0.0139 & 0.0461 & 0.1494 & \textbf{0.0149} & \textbf{0.0494} & \textbf{0.1568} & \textbf{0.0149} & \textbf{0.0475} & \textbf{0.1522} \\
\bottomrule
\end{tabular}
\end{table*}

Table~\ref{tab:time} reports mean computation times for $M=200$. MPCA requires about $0.19$--$0.20$ seconds and SMPCA about $0.81$--$0.83$ seconds per run. The additional cost arises mainly from the spatial-median calculation and sign normalization. Under the same implementation and hardware, ROMPCA requires approximately $277$--$345$ seconds and the TPCA-\(L_p\) variants approximately $37$--$73$ seconds. These timings explain why the 1000-replication experiment is limited to MPCA and SMPCA.

\begin{table}[htbp]
\centering
\footnotesize
\setlength{\tabcolsep}{4pt}
\caption{Mean computation time in seconds for $M=200$}
\label{tab:time}
\begin{tabular}{@{}lccc@{}}
\toprule
Method & Normal & $t_{2.5}$ & Mixture \\ 
\midrule
MPCA      & 0.1954   & 0.1990   & 0.1941   \\
SMPCA     & 0.8313   & 0.8132   & 0.8154   \\
ROMPCA    & 276.8402 & 345.2287 & 278.3828 \\
TPCA-\(L_{0.5}\) & 70.7944  & 73.0293  & 72.7211  \\
TPCA-\(L_1\)   & 39.2991  & 37.2808  & 37.6484  \\
TPCA-\(L_{1.5}\) & 60.0911  & 58.1229  & 59.0182  \\
\bottomrule
\end{tabular}
\end{table}

\subsection{Dimension selection}
\begin{figure*}[t]
\centering
\includegraphics[width=\textwidth]{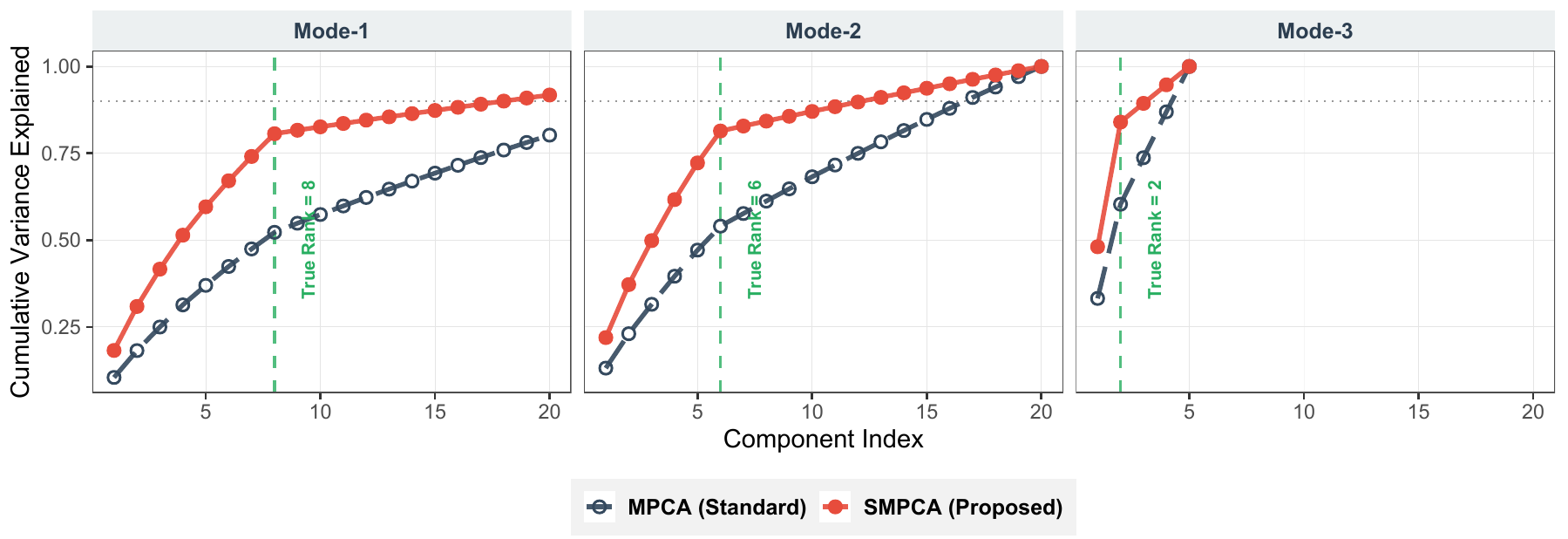}
\caption{Cumulative contribution curves under centered Gaussian-mixture noise with baseline variance $0.1$, variance-inflation factor $9$ and contamination probability $\pi_{\mathrm{out}}=0.2$. Vertical dashed lines mark the data-generating multilinear ranks $(8,6,2)$}
\label{fig:rank_selection}
\end{figure*}

We next compare cumulative-contribution curves under the Gaussian-mixture noise used above. The contamination probability is $\pi_{\mathrm{out}}=0.2$. Figure~\ref{fig:rank_selection} contrasts the mode-wise curves formed from the standard MPCA scatter matrices and the SMPCA spatial-sign covariance matrices.

Under mixture contamination, the MPCA curves are smoothed by variance inflation and the boundaries between signal and noise directions are less distinct. The SMPCA curves display substantially clearer empirical elbows near the data-generating ranks $(8,6,2)$. This behavior is consistent with the perturbation result in Theorem~\ref{thm:rank-consistency}; exact recovery by a fixed cumulative threshold nevertheless requires the population margin in \eqref{eq:rank-margin}.

\section{Face-image analysis}
We assess reconstruction performance on the Georgia Tech face database \citep{georgiatechfacedb}. The database contains 15 color images for each of 50 individuals. We compare SMPCA with MPCA, ROMPCA \citep{hirari2026robust} and TPCA-\(L_p\) \citep{7924765} under controlled Gaussian-mixture contamination.

\subsection{Experimental design}
For each parameter configuration, we perform five independent repetitions. In every repetition, we draw 10 individuals at random and use all available images for those individuals. To make all methods computationally feasible under a common resolution, each color image is downsampled by a factor of $0.2$ in both spatial dimensions.

\begin{figure*}[t]
    \centering
    \includegraphics[width=0.9\textwidth]{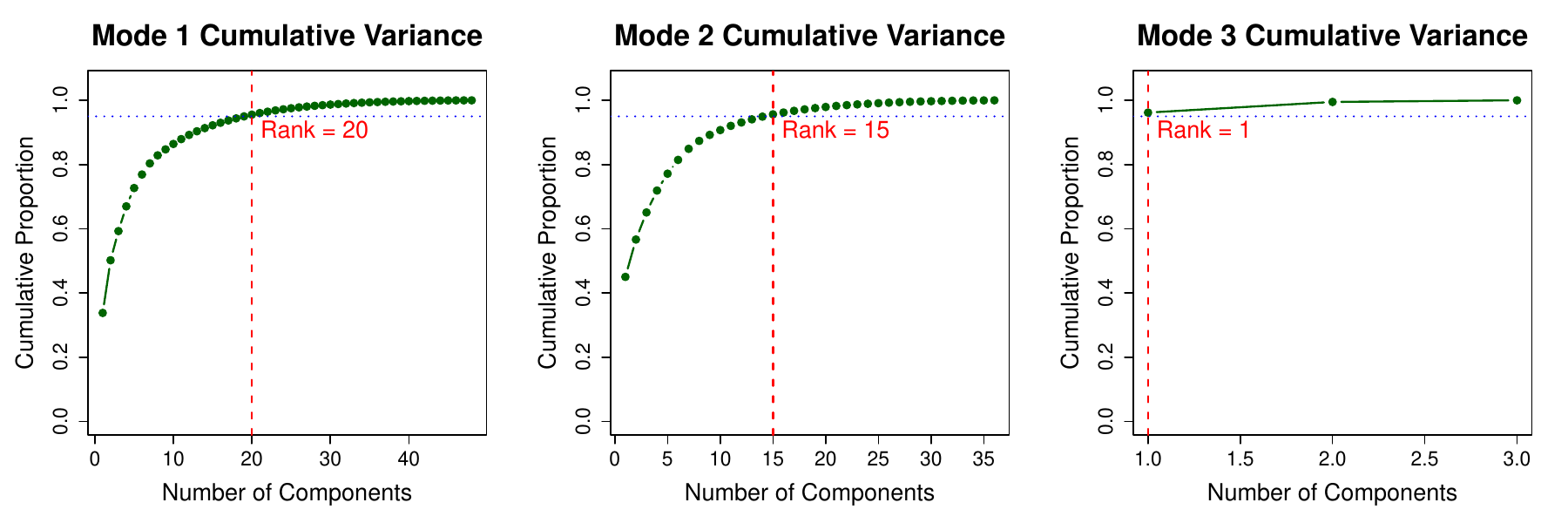}
    \caption{SMPCA dimension selection from the eigenvalues of the mode-wise spatial-sign covariance matrices. The selected dimensions corresponding to a 95\% cumulative contribution threshold are $(20, 15, 1)$ at scale $0.2$}
    \label{fig:rank_selection_face}
\end{figure*}

Independent Gaussian-mixture noise is added entrywise, with baseline standard deviation $\sigma\in\{0.05,0.10\}$, contamination probability $\pi_{\mathrm{out}}\in\{0.2,0.4\}$ and outlier scale multiplier $a_{\mathrm{out}}\in\{3,5,8,10\}$. The contaminated pixel values are clipped to $[0,1]$. Using a cumulative-contribution threshold \(\tau=0.95\), SMPCA selects mode dimensions \((20,15,1)\) at image scale $0.2$; Figure~\ref{fig:rank_selection_face} displays the corresponding curves. These dimensions are used for every method so that the reconstruction comparison is not confounded by different retained dimensions.

\subsection{Evaluation criteria}
Let $\mathcal X_m^{\mathrm{cl}}$ denote the uncontaminated image tensor and $\widehat{\mathcal X}_m^{\mathrm{rec}}$ its reconstruction. We report four criteria.
\begin{enumerate}[label=(\alph*)]
\item The average relative projection error is
\[
\operatorname{ARPE}
=\frac1M\sum_{m=1}^M
\frac{\|\mathcal X_m^{\mathrm{cl}}-\widehat{\mathcal X}_m^{\mathrm{rec}}\|_F}{\|\mathcal X_m^{\mathrm{cl}}\|_F}.
\]
Smaller values indicate better recovery.
\item With pixel intensities scaled to $[0,1]$, the mean peak signal-to-noise ratio is
\[
\operatorname{PSNR}
=\frac1M\sum_{m=1}^M10\log_{10}(\operatorname{MSE}_m^{-1}),
\]
where, with $p_{\mathrm{img}}$ denoting the number of scalar entries in an image tensor,
$\operatorname{MSE}_m=p_{\mathrm{img}}^{-1}\|\mathcal X_m^{\mathrm{cl}}-\widehat{\mathcal X}_m^{\mathrm{rec}}\|_F^2$.
\item The structural similarity index is averaged over image windows. For two local windows $x$ and $y$,
\[
\begin{aligned}
\operatorname{SSIM}(x,y)
&=\frac{2\mu_x\mu_y+c_{\mathrm S,1}}
{\mu_x^2+\mu_y^2+c_{\mathrm S,1}}\\
&\quad\times
\frac{2\sigma_{xy}+c_{\mathrm S,2}}
{\sigma_x^2+\sigma_y^2+c_{\mathrm S,2}}.
\end{aligned}
\]
Here $\mu_x$ and $\mu_y$ are the local means, $\sigma_x^2$ and $\sigma_y^2$ are the local variances, $\sigma_{xy}$ is the local covariance, and $c_{\mathrm S,1},c_{\mathrm S,2}>0$ are fixed stabilizing constants. Larger PSNR and SSIM values indicate better reconstruction.
\item Computation time is the elapsed time, in seconds, required for subspace estimation under the reported implementation and hardware.
\end{enumerate}

\subsection{Results}
\begin{figure*}[t]
\centering
\begin{minipage}[t]{0.46\textwidth}
\centering
\includegraphics[width=\linewidth]{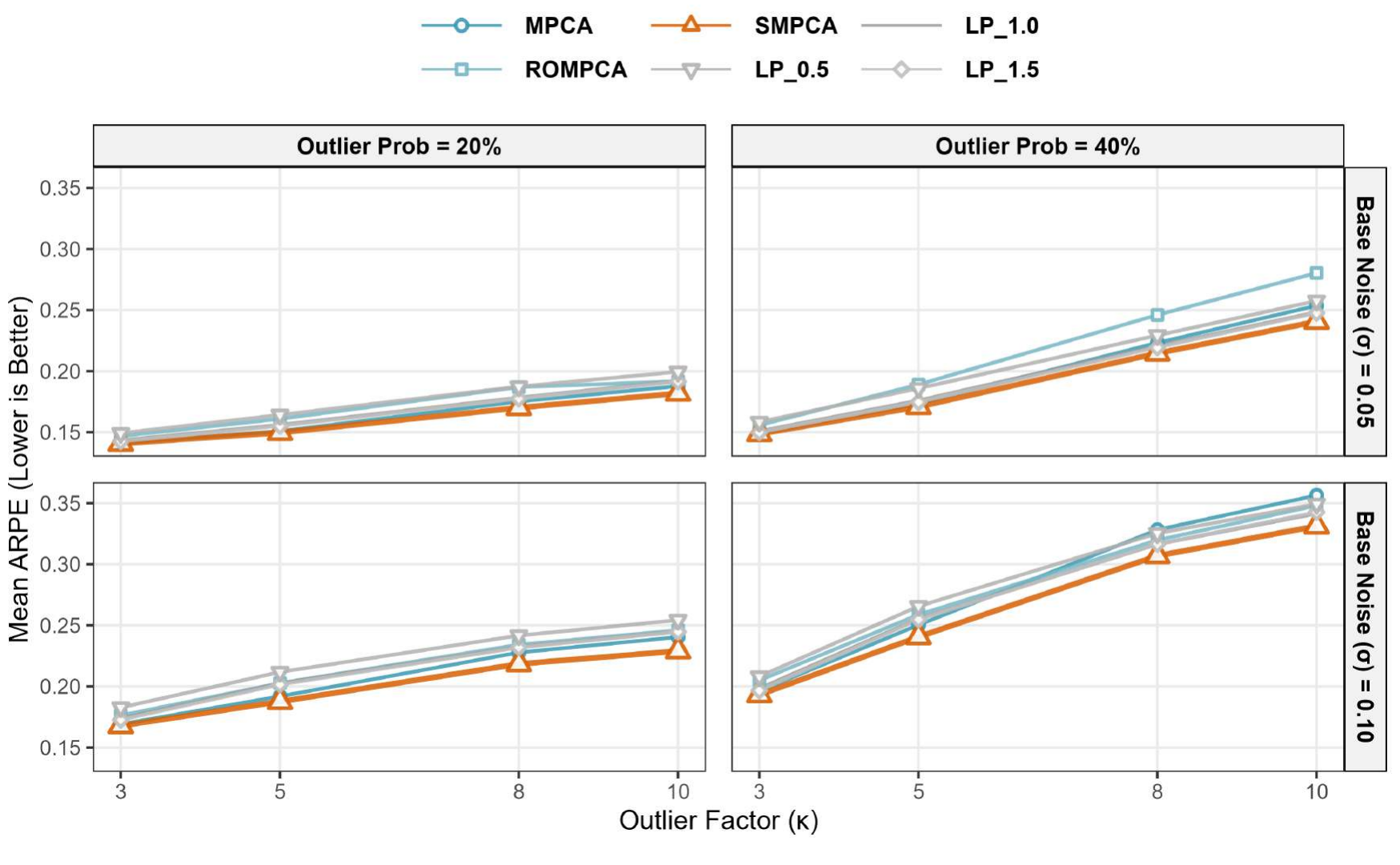}\\[-1mm]
\textbf{(a)} ARPE (smaller is better)
\end{minipage}\hfill
\begin{minipage}[t]{0.46\textwidth}
\centering
\includegraphics[width=\linewidth]{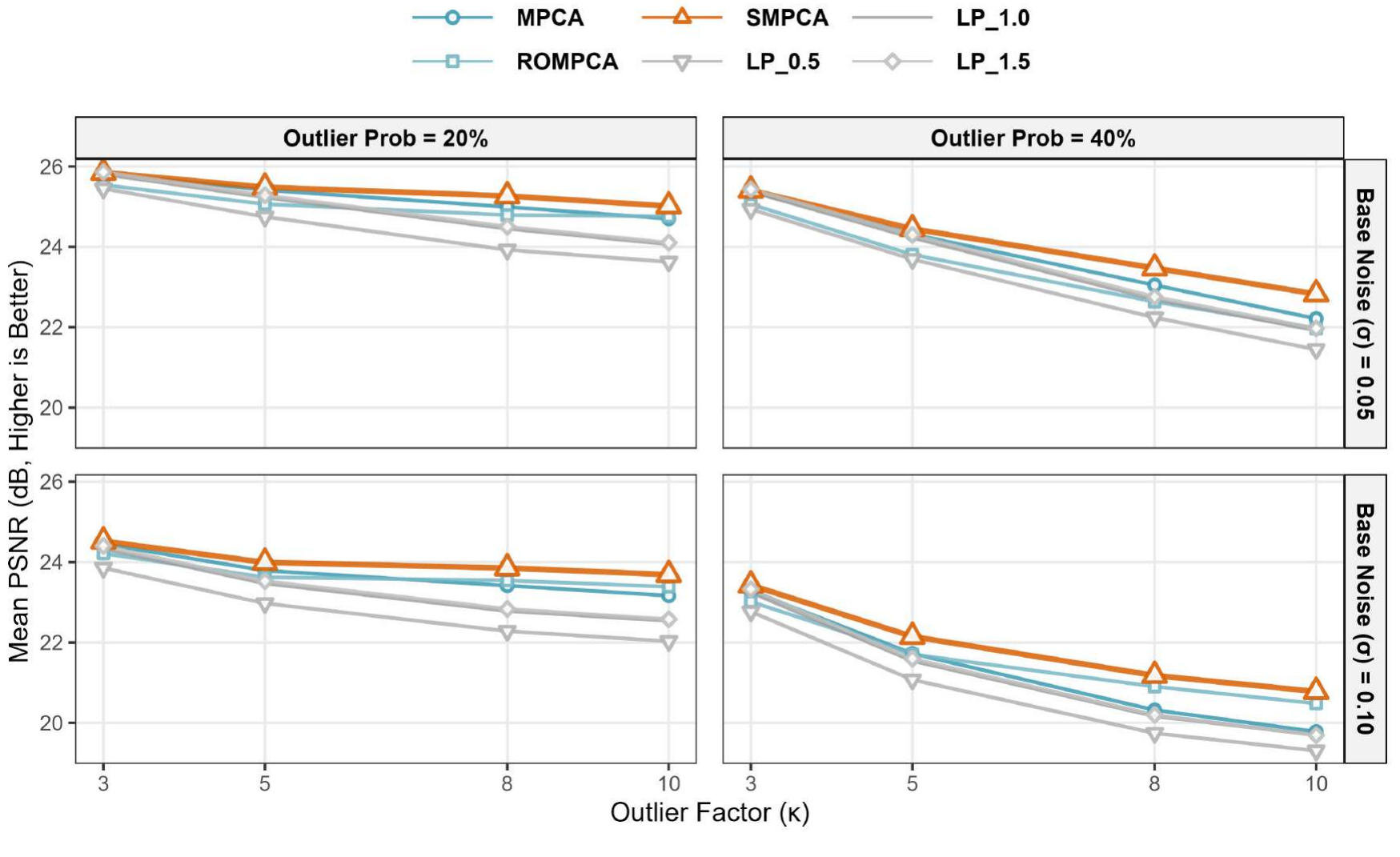}\\[-1mm]
\textbf{(b)} PSNR (larger is better)
\end{minipage}

\vspace{2mm}
\begin{minipage}[t]{0.46\textwidth}
\centering
\includegraphics[width=\linewidth]{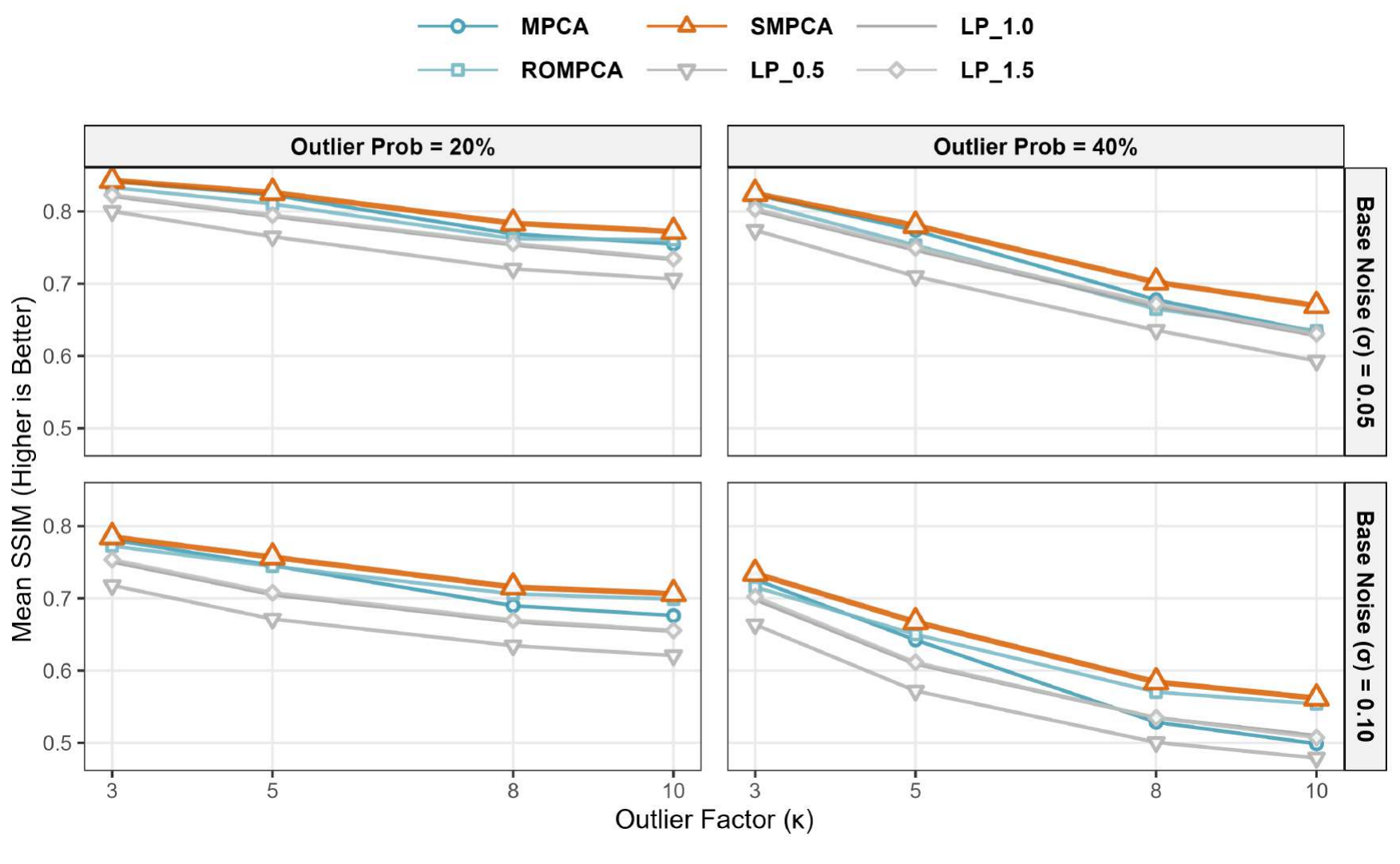}\\[-1mm]
\textbf{(c)} SSIM (larger is better)
\end{minipage}\hfill
\begin{minipage}[t]{0.46\textwidth}
\centering
\includegraphics[width=\linewidth]{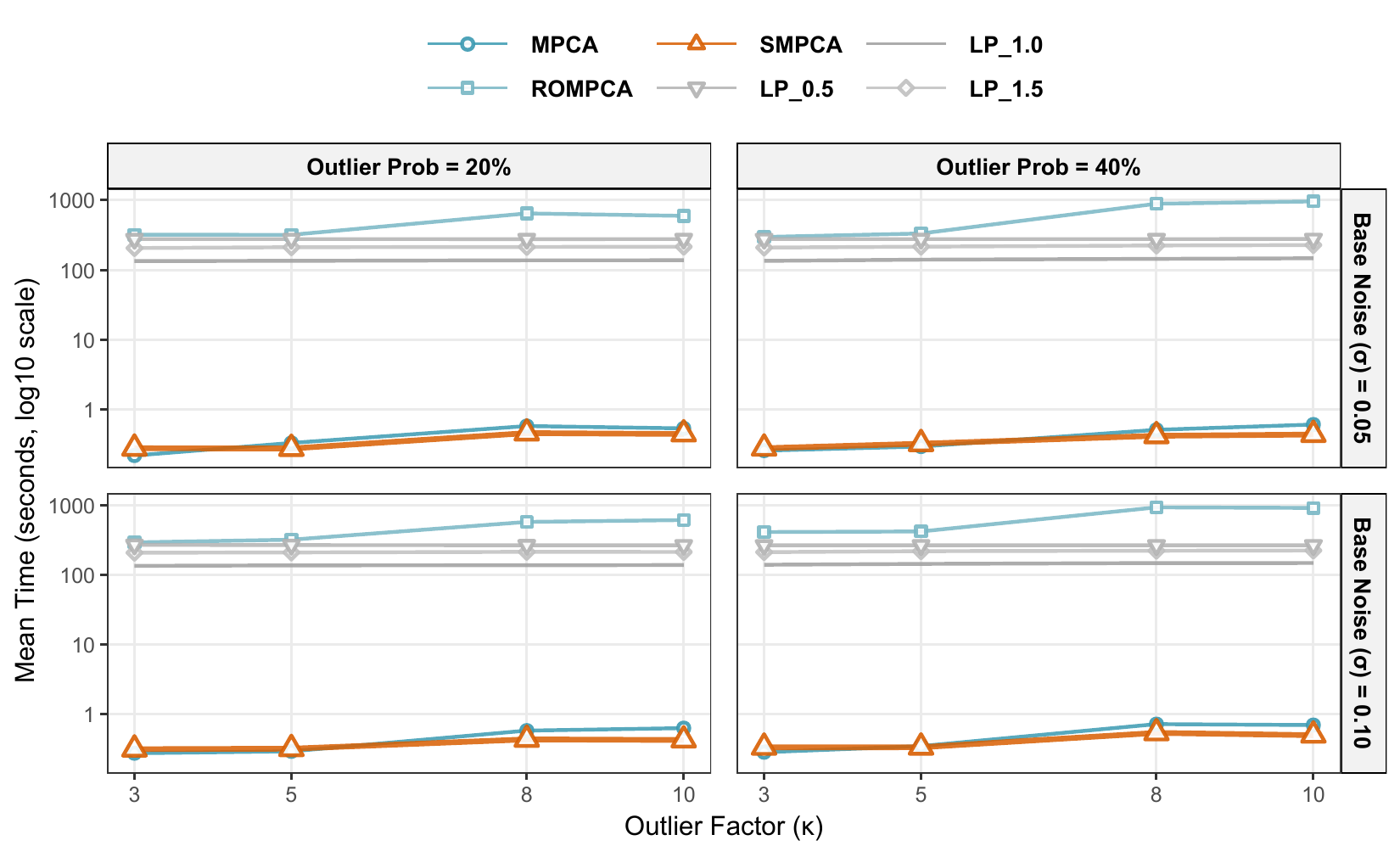}\\[-1mm]
\textbf{(d)} Computation time ($\log_{10}$ scale)
\end{minipage}
\caption{Reconstruction accuracy and computation time under Gaussian-mixture contamination at image scale 0.2}
\label{fig:robustness}
\end{figure*}

Figure~\ref{fig:robustness}(a)--(c) reports ARPE, PSNR and SSIM\@. MPCA deteriorates as either the contamination probability or the outlier scale increases. SMPCA is considerably more stable because the spatial-sign transformation maps every centered tensor to a tensor with bounded Frobenius norm. Across the reported configurations, SMPCA is either the best-performing method or close to it; it generally matches ROMPCA and improves on the TPCA-\(L_p\) variants.

Figure~\ref{fig:robustness}(d) reports mean computation time on a logarithmic scale. SMPCA completes each analysis in less than one second in this experiment. ROMPCA and the TPCA-\(L_p\) procedures require tens to hundreds of seconds. Thus the spatial-median and sign preprocessing adds a modest cost relative to MPCA but remains substantially less expensive than the alternative robust procedures considered here.

\begin{figure*}[t]
    \centering
    \includegraphics[width=0.8\textwidth]{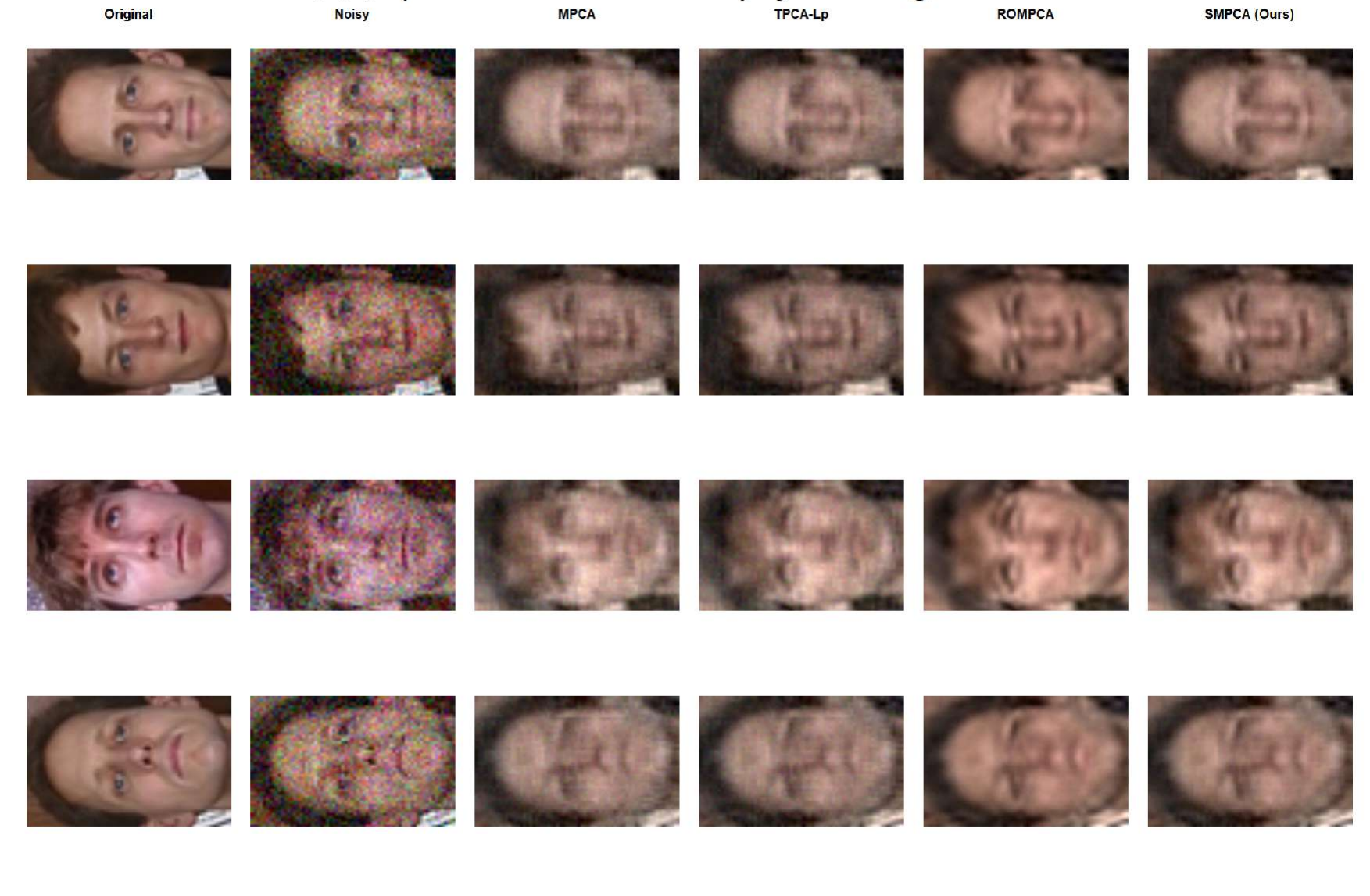}
    \caption{Representative face reconstructions. From left to right: clean image, contaminated image, MPCA, TPCA-\(L_p\), ROMPCA and SMPCA. Source: Georgia Tech Face Database \citep{georgiatechfacedb}}
    \label{fig:face_reconstruction_visual}
\end{figure*}

The representative reconstructions in Figure~\ref{fig:face_reconstruction_visual} agree with the numerical summaries. MPCA retains visible contamination artifacts, whereas ROMPCA and SMPCA yield visibly cleaner reconstructions. In the reported implementation, SMPCA achieves comparable or better reconstruction accuracy with substantially lower computation time than ROMPCA.

\section{Discussion}
We have proposed SMPCA, a robust multilinear dimension-reduction procedure based on spatial-median centering, tensor spatial signs and alternating mode-wise eigendecompositions. Under a separable tensor elliptical model, the population block scatters identify the mode scale eigenspaces through a weighted marginalization of the product-basis eigenvalues of the spatial-sign covariance matrix. This representation also clarifies the role of tensor structure. SMPCA searches over Kronecker-structured subspaces parameterized by a product of mode-wise Grassmann manifolds, whereas vectorized PCA and vectorized spatial-sign PCA search over unrestricted rank-\(K\) subspaces. Proposition~\ref{prop:rectangular-target} characterizes exactly when the two targets coincide and shows that the mode loading spaces remain identifiable when they do not. The Grassmann dimensions displayed in Section~\ref{sec:target-comparison} quantify the resulting reduction in structural degrees of freedom.

The main statistical results give explicit rates for the initial loading spaces, every exact block update, the joint Kronecker projector and the induced reconstruction map. We also establish consistency of cumulative-contribution dimension selection and convergence of the alternating objective values. The auxiliary theory shows why the natural location rate $\sqrt{p/M}$, obtained under assumptions on $R/\sqrt p$, produces only an $M^{-1/2}$ contribution to the spatial-sign covariance error.

The numerical studies indicate that SMPCA incurs little loss relative to MPCA under Gaussian sampling and is substantially more stable under heavy-tailed and mixture contamination. In the reported implementations, the additional preprocessing cost of SMPCA relative to MPCA remains small compared with the computation times of ROMPCA and TPCA-\(L_p\).

Several extensions merit further study. The operator-norm bounds used here pass through the full $p\times p$ spatial-sign covariance matrix; sharper rates may be obtainable by exploiting the Kronecker structure directly. Inference for mode eigenvalues and loading projectors would complement the present estimation theory. Other directions include online, supervised and non-negative variants, and robust extensions of CANDECOMP/PARAFAC decompositions \citep{Harshman1970FoundationsOT,Carroll_Chang_1970}.

\backmatter

\section*{Statements and Declarations}

\noindent\textbf{Competing interests.} The authors have no relevant financial or non-financial interests to disclose.\par
\noindent\textbf{Data availability.} The Georgia Tech Face Database analyzed in Section~6 is publicly available at \url{https://www.anefian.com/research/face_reco.htm} and is cited in the reference list \citep{georgiatechfacedb}. All remaining numerical results are based on simulated data generated from the models and parameter settings reported in Section~5.\par
\noindent\textbf{Materials availability.} Not applicable.\par
\noindent\textbf{Code availability.} The implementation used for the numerical studies is available from the corresponding author on reasonable request.\par
\noindent\textbf{Author contributions.} All authors contributed to the conception and methodology of the study, analysis and interpretation of the results, and preparation of the manuscript. All authors read and approved the final manuscript.

\setlength{\bibsep}{0.10em}
\bibliography{ref}

\end{document}